\documentclass[12pt]{article}

\usepackage[a4paper,total={18cm,27cm}]{geometry}

\usepackage[utf8]{inputenc}
\usepackage{amsmath}
\usepackage{amsfonts}

\usepackage{graphicx}
\usepackage{array}
\usepackage{caption}

\usepackage{xcolor}
\colorlet{shadecolor}{yellow!20}

\usepackage{authblk}

\usepackage{hyperref}

\usepackage{comment}   

\newcommand{\NI}{\vspace{0.2cm}\noindent}

\newcommand{\is}{\!=\!}

\begin{document}


\title{Structural and dynamical strategies to prevent\\ runaway excitation in reservoir computing}


\author[1]{Claus Metzner}
\author[1,2,3,4]{Achim Schilling}
\author[1]{Andreas Maier}
\author[2]{Thomas Kinfe}
\author[1,2,3,4]{Patrick Krauss}

\affil[1]{\small Cognitive Computational Neuroscience Group, Pattern Recognition Lab, Friedrich-Alexander-University Erlangen-Nürnberg (FAU), Germany}
\affil[2]{\small Neuromodulation and Neuroprosthetics, University Hospital Mannheim, University Heidelberg, Germany}
\affil[3]{\small Neuroscience Lab, University Hospital Erlangen, Germany}
\affil[4]{\small BGU Ludwigshafen, Germany}

\maketitle


\begin{abstract}
\large
\NI Reservoirs, typically implemented as recurrent neural networks with fixed random connection weights, can be combined with a simple trained readout layer to perform a wide range of computational tasks. However, increasing the magnitude of reservoir connection weights to exploit nonlinear dynamics can cause the network to develop strong spontaneous activity that drives neurons into saturation, dramatically degrading performance. In this work, we investigate two distinct countermeasures against such runaway excitation. The first approach introduces a subtle non-homogeneous structure into the matrix of connection weigths $w_{ij}$, without altering the overall probability distribution $p(w)$. We identify several favorable structuring principles, such as creating a small subset of neurons with weaker-than-average input connections. Even if the rest of the reservoir falls into runaway saturating behavior, this weakly coupled subset remains in a mildly nonlinear regime whose dynamics can still be exploited by the readout layer. The second approach implements a form of automatic gain control, in which a dedicated control unit dynamically regulates the reservoir's average global activation toward an optimal setpoint. Although the control unit modulates the excitability of the reservoir only via a global gain factor, this mechanism substantially enlarges the dynamical regime favorable for computation and renders performance largely independent of the underlying connection statistics.
\end{abstract}

\newpage

\section{Introduction}

\NI Deep learning has made substantial progress in recent years \cite{lecun2015deep,alzubaidi2021review}, partly propelled by the emergence of large language models \cite{min2023recent}. These models primarily depend on feedforward architectures, where information flows unidirectionally from input to output. Recurrent neural networks (RNNs), in contrast, feature feedback connections that enable them to function as self-sustaining dynamical systems \cite{maheswaranathan2019universality}, maintaining internal activity without requiring continuous external input.

\NI Several universal capabilities characterize RNNs, including their theoretical capacity to approximate arbitrary mappings \cite{schafer2006recurrent} and replicate general dynamical systems \cite{aguiar2023universal}. These properties, along with their practical advantages, have generated sustained interest in understanding the internal mechanisms of these networks. RNNs can integrate information over prolonged temporal intervals \cite{jaeger2001echo,schuecker2018optimal,busing2010connectivity,dambre2012information,wallace2013randomly,gonon2021fading}, and they possess the ability to construct compact yet rich internal representations through a balance of dimensional compression and expansion \cite{farrell2022gradient}.

\NI Research has also examined how RNN dynamics can be controlled and stabilized, especially when subjected to noise from both internal and external sources \cite{rajan2010stimulus,jaeger2014controlling,haviv2019understanding,molgedey1992suppressing,ikemoto2018noise,krauss2019recurrence,bonsel2021control,metzner2022dynamics}. RNNs have additionally attracted interest as computational frameworks for brain-like information processing \cite{barak2017recurrent}. Sparse RNNs with limited connections per unit, which mirror the structural organization of biological neural circuits \cite{song2005highly}, have demonstrated benefits for memory retention and representational efficiency \cite{brunel2016cortical,narang2017exploring,gerum2020sparsity,folli2018effect}.

\NI Our earlier work established a systematic investigation into how structural features influence RNN dynamics, beginning with small three-neuron motifs \cite{krauss2019analysis}. Extending this foundation, we showed how large autonomous networks can be controlled through three critical parameters: weight distribution width $w$, connection density $d$, and excitation/inhibition balance $b$ \cite{krauss2019weight,metzner2022dynamics}. We further explored how these systems respond to noise, uncovering resonance-like phenomena and stochastic stabilization effects \cite{bonsel2021control,schilling2022intrinsic,krauss2016stochastic,krauss2019recurrence,schilling2021stochastic,schilling2023predictive,metzner2024recurrence}.

\NI Our more recent research has investigated how particular dynamical characteristics and structural regularities affect reservoir computing system performance.

\NI In \cite{metzner2025nonlinear}, we systematically manipulated the degree of neuronal nonlinearity and the intensity of recurrent coupling to examine their effects on classification accuracy in synthetic tasks. Remarkably, even 'quiet' reservoirs exhibiting minimal internal dynamics and extremely weak nonlinearity demonstrated the capacity to produce linearly separable representations for the readout layer, provided that subtle high-dimensional state features were leveraged. Although task performance generally declined under highly chaotic dynamics, accuracy frequently peaked near transition regions between chaotic and oscillatory or fixed-point regimes, providing support for the "edge of chaos" hypothesis.

\NI In \cite{metzner2025organizational}, we examined how biologically motivated structural regularities influence both the autonomous dynamics of RNNs and their functional performance in reservoir computing. These regularities included Dale's principle, the creation of stronger and weaker patches in the connection weight matrix to create local heterogeneity \cite{sporns2016modular,meunier2010modular}, as well as reciprocal symmetry similar to Hopfield networks \cite{hopfield1982neural}. Our findings indicated that Dale-conforming output weights and heterogeneous connectivity patches tend to improve task accuracy, whereas strong reciprocal symmetry proved detrimental, frequently pushing the system toward premature saturation and diminishing its ability to discriminate between inputs. Collectively, these results already hint at potential advantages of incorporating organizational structure into randomly connected networks.

\NI In the present work, we focus on the problem of runaway excitations in strongly connected reservoir networks, which can arise independently of the present input. These excitations drive neurons deep into the saturation of their sigmoidal activation functions and manifest at the network level as global synchronous oscillations, chaotic fluctuations, or global fixed points. In any of these cases, the computational performance of the reservoir computer typically degrades dramatically because the dynamics is no longer predominantly determined by the present input. To address this problem, we test two different approaches.

\NI Our first approach is to start with a statistically homogeneous connection weight matrix and then to re-distribute its elements in order to create heterogeneous structure in the matrix, yet without altering the global probability density distribution of the elements. The idea is that runaway excitations can often be interpreted as a herding effect driven by strong negative or positive feedback between all neurons in the reservoir. By creating diverse subsets of neurons with different statistical properties, such global herding effects may be suppressed.

\NI Our second approach is a direct dynamical regulation of network activity. We compute a running average of the overall network activation level, measured as the root-mean-squared activation of all reservoir neurons. A control unit then regulates a global prefactor of the weight matrix to keep the activity close to a setpoint that has been optimized for computation. Although highly simplified, this mechanism bears a conceptual resemblance to global regulatory processes in biological neural systems, where overall excitability and effective coupling strength are modulated by neuromodulators, hormones, or glial interactions. While our model does not attempt to capture such mechanisms in detail, it illustrates how a coarse-grained, global control signal can stabilize network dynamics and expand the regime of useful computation.

\section{Methods}

\subsection{Generation of a Homogeneous Weight Matrix}
\label{GenHomWeiMat}

\NI The weight matrix $\mathbf{W}$ of the reservoir's recurrent connections is random but controlled by three statistical parameters: the {\bf density} $d$ of non-zero connections, the excitatory/inhibitory {\bf balance} $b$, and the recurrent {\bf coupling strength} $w$, which is defined by the standard deviation of the Gaussian distribution of weight magnitudes.  

\NI The density $d$ ranges from $d = 0$ (isolated neurons) to $d = 1$ (fully connected network). The balance $b$ ranges from $b = -1$ (purely inhibitory connections) to $b = +1$ (purely excitatory connections), with $b = 0$ corresponding to a perfectly balanced system. The coupling strength $w$ can take any positive value, where $w = 0$ corresponds to unconnected reservoir neurons.  

\NI The bias terms $\mathbf{b_w}$ applied to the reservoir neurons are also drawn from a Gaussian distribution with zero mean and a fixed variance, ensuring that the reservoir operates in a dynamic regime suitable for information processing.

\NI In order to construct an $N \times N$ weight matrix with given parameters $(b, d, w)$, we proceed as follows:  

\NI First, we generate a matrix $\mathbf{M}^{(magn)}$ of weight magnitudes by drawing the $N^2$ matrix elements independently from a zero-mean normal distribution with standard deviation $w$ and then taking the absolute value.  

\NI Next, we generate a random binary matrix $\mathbf{B}^{(nonz)} \in \{0,1\}^{N \times N}$, where the probability of a matrix element being $1$ is given by the density parameter $d$, that is, $p_1 = d$.  

\NI We then generate another random binary matrix $\mathbf{B}^{(sign)} \in \{-1, +1\}^{N \times N}$, in which the probability of a matrix element being $+1$ is given by  
\begin{equation}
p_{+1} = \frac{1 + b}{2}
\end{equation}  
where $b$ is the balance parameter.  

\NI Finally, the weight matrix is constructed by element-wise multiplication:  
\begin{equation}
W_{mn} = M^{(magn)}_{mn} \cdot B^{(nonz)}_{mn} \cdot B^{(sign)}_{mn}
\end{equation}  

\NI Note that throughout this paper, the density parameter is always set to the maximum value of $d = 1$.

\NI The biases $b_{w,n}$ of the reservoir are drawn independently from a zero-mean normal distribution with a standard deviation of $w'$. Note that throughout this paper, the strength of the reservoir bias is always set to $w' = 0.1$. Thus, even for an uncoupled reservoir with $w = 0$ and zero input, the neurons will have non-zero resting values due to the random biases. 

\subsection{Generation of a Structured Weight Matrix}

\NI Starting from a statistically homogeneous weight matrix $\mathbf{W}$ of size $N \times N$, generated under fixed parameters width $w$, density $d$, and balance $b$, our goal is to introduce controlled heterogeneity and structural organization without altering the overall distribution of weight values. To this end, we apply a three-step procedure consisting of mask generation, element sorting, and structured reassembly.

\NI In the first step, a binary mask of the same shape as $\mathbf{W}$ is constructed. The mask entries take values 0 or 1, and their spatial arrangement is determined by a mode parameter $\texttt{mskMod}$. In \emph{random} mode, a fraction $\texttt{oneFrc}$ of all matrix elements is selected uniformly at random and marked with 1. In \emph{rows} mode, a fraction $\texttt{oneFrc}$ of all rows is randomly selected and entirely marked with 1. In \emph{cols} mode, a fraction $\texttt{oneFrc}$ of all columns is randomly selected and entirely marked with 1. In \emph{diagBlocks} mode, the mask consists of a chain of contiguous $S \times S$ blocks of ones arranged along the main diagonal; this requires $\mathbf{W}$ to be square and its dimension to be divisible by $S$. In this latter case, the parameter $\texttt{oneFrc}$ is not used.

\NI In the second step, all elements of $\mathbf{W}$ are collected into a one-dimensional list and sorted according to a mode parameter $\texttt{srtMode}$. Sorting can be performed with respect to the raw weight values or their absolute magnitudes, each in either ascending or descending order. This step establishes a controlled ranking of weights while preserving the original multiset of values and hence the distribution implied by $w$, $d$, and $b$.

\NI In the final step, the sorted list of weight values is redistributed over the matrix positions according to the previously generated mask. The number of elements assigned to mask entries equal to 1 and 0 is determined by their respective counts in the mask. The sorted list is split accordingly into two subsets, which are randomly permuted within themselves to avoid introducing unintended fine-scale order. The shuffled subsets are then placed into the matrix positions corresponding to mask entries 1 and 0, respectively.

\NI The overall procedure thus preserves the global weight distribution exactly, while introducing structured heterogeneity at the level specified by $\texttt{mskMod}$, $\texttt{oneFrc}$, $S$, and $\texttt{srtMode}$. This allows us to generate systematically structured weight matrices from an initially homogeneous ensemble without modifying the underlying statistics of the weights.

\subsection{Design of Reservoir Computer (RC)}

\NI The RC consists of an input layer, a recurrent reservoir, a readout layer and optionally, a control unit. The input data comprises $E$ consecutive episodes, each representing, for example, a pattern to be classified, or a priming sequence triggering the output of a specific target response. This data stream is fed into the reservoir and circulates through the system while propagating toward the output.

\NI At each time step $t$, the input layer receives $M$ parallel signals $x_m^{(t)} \in [-1, +1]$. These are linearly transformed by the input matrix $\mathbf{I}$ of size $N \times M$ and injected into the reservoir, as described by Eq.~\ref{yEq}. Each input episode spans $T$ time steps.

\NI The {\bf input layer} consists solely of the matrix $\mathbf{I}$ and is therefore purely linear. Its elements $I_{mn}$ are drawn independently from a normal distribution with zero mean and standard deviation $w_I$. To study autonomous RNN dynamics, we set $w_I = 0$, effectively decoupling the reservoir from external input.

\NI The {\bf reservoir} comprises $N$ recurrently connected neurons with $\tanh$ activation. At each time step, all neuron states $y_n$ are updated in parallel. Each neuron receives a bias term $b_{w,n}$, input from the external signals via $\mathbf{I}$, and recurrent input via the weight matrix $\mathbf{W}$ (see Eq.~\ref{yEq}), which can be dynamically modulated by a gain factor $g$ from the control unit. In settings without control, we set $g=1$. Initial states $y_n^{(0)}$ are drawn uniformly from $[-1, +1]$ and kept fixed for repeated simulations of the same reservoir. Different weight matrices receive independent initial states.

\NI The {\bf readout layer} performs an affine-linear transformation of the reservoir states $y_n$ using a $K \times N$ output matrix $\mathbf{O}$ and a bias vector $\mathbf{b_o}$, as in Eq.~\ref{zEq}. These parameters are trained via the pseudoinverse method (see below).

\NI In the sequence generation task, which serves as the main information processing benchmark in this work, the continuous outputs $z_k$ of the readout layer directly form the vectors of the output sequence. In classification tasks, by contrast, the $z_k$ represent soft votes that are converted into discrete predicted class labels $c$ using the argmax function. This final step introduces a nonlinearity that sharpens the class boundaries in the output space.

\NI In summary, the RC is governed by the following equations:

\begin{eqnarray}
y_n^{(t)} & = & \tanh \left( 
  b_{w,n} + \sum_m I_{nm} x_m^{(t-1)} + 
  g^{(t)}\;\sum_{n^{\prime}} W_{nn^{\prime}} y_{n^{\prime}}^{(t-1)} 
\right) \label{yEq} \\
z_k^{(t)} & = & b_{o,k} + \sum_n O_{kn} y_n^{(t)} \label{zEq} \\
c^{(t)} & = & \operatorname*{arg\,max} \left\{ z_k^{(t)} \right\}
\end{eqnarray}

\subsection{Unit for Automatic Gain Control (AGC)}

\NI The control unit computes a running average $\overline{A}^{(t)}$ of the reservoir's momentary root-mean-square activation $A^{(t)}$:

\begin{eqnarray}
A^{(t)} & = & \sqrt{\frac{1}{N}\sum_n \left(y_n^{(t)}\right)^2} \label{aEq} \\
\overline{A}^{(t)} & = & \mu A^{(t)} + (1\!-\!\mu)\overline{A}^{(t\!-\!1)} \label{aBarEq} \\
\end{eqnarray}

\NI At time zero, we set $\overline{A}^{(0)}=0$. The {\bf mixing factor} $\mu$, ranging between 0 and 1, controls how strongly the present network activation is weighted compared to the past average. 

\NI The running average $\overline{A}^{(t)}$ is compared to an {\bf activity setpoint $\alpha$}, which in our system has a range from 0 to 1. Depending on whether the difference is positive or negative, the current gain factor $g$ is multiplied with an exponential function that is smaller or larger than one, respectively:

\begin{equation}
g^{(t)} = g^{(t\!-\!1)} \exp{\left( -\epsilon\; (\overline{A}^{(t)}-\alpha) \right)}
\end{equation}

\NI At time zero, we set $\overline{g}^{(0)}=1$. The small positive {\bf feedback parameter} $\epsilon$ determines the sensitivity of the gain control mechanism. In the special case $\epsilon=0$, the automatic gain control is effectively switched off. 

\NI The results with AGC shown in this paper were based on the following parameter settings: $\mu=0.1$, $\alpha=0.25$ and $\epsilon=0.25$.

\subsection{Sequence Generation Task}

\NI In this task, the reservoir computer functions as a deterministic system that maps an input sequence $X$ of real-valued vectors onto a corresponding output sequence $Z$:
\[
X \in [-1,+1]^{TI \times M} \;\longrightarrow\; Z \in [-1,+1]^{TO \times K}
\]
Here, $TI$ and $TO$ denote the temporal lengths of the input and output sequences, and $M$ and $K$ their respective vector dimensions.

\NI At the beginning of each episode, a randomly chosen input sequence $X$ from one of $N_{DC}$ discrete classes is fed into the network, driving it into a class-specific internal configuration or ‘priming state’. After the input ends, the reservoir evolves autonomously through a sequence of internal states $Y$, which the linear readout layer is trained to map onto the corresponding target sequence $Z$.

\NI Since the system is strictly deterministic, it will always produce the same trajectory for each distinct priming state. Provided that the induced trajectory is not a cyclic attractor with a period shorter than $TO$, the readout should in principle be able to convert the trajectory into the desired output sequence.

\NI However, because the reservoir is not reset at the beginning of each episode, residues of previous states may persist, such that the priming state is not exactly identical every time a given class is selected. If the reservoir operates in a chaotic regime, these small differences can be amplified, producing an effectively unpredictable trajectory that cannot be mapped to the correct target.

\NI In practice, the target mapping also fails if the reservoir neurons enter the saturated ‘digital’ state, since the resulting trajectory is insufficiently rich and lacks the necessary high-dimensional diversity.

\NI Successful performance therefore requires a balance between stability and richness: the reservoir must forget prior excitation rapidly enough to respond reproducibly to identical inputs, while still maintaining sufficient temporal and spatial diversity across neurons.

\NI To systematically explore the influence of network parameters under controlled conditions, we employ a minimal task configuration with $M\!=\!K\!=\!2$, $TI\!=\!1$, $TO\!=\!2$, and $N_{DC}\!=\!2$. The class-specific input sequences $X_c$ and their target sequences $Z_c$ are predefined, with all vector components drawn independently from a uniform distribution over $[-1,+1]$. In each episode, one of the $N_{DC}$ cases is selected at random.

\subsection{Optimal Readout Layer Using Pseudoinverse}

\NI The optimal weights and biases of the readout layer can be efficiently computed with the method of the pseudoinverse, based on the sequence of reservoir states and the target output (Compare, for example, section 3.4. in \cite{cucchi2022hands}). Following these ideas, we proceed as follows:  

\NI Let \( Y \in \mathbb{R}^{(E-1) \times N} \) be the matrix of reservoir states directly after each input episode, where \( E \) is the total number of episodes and \( N \) is the number of reservoir neurons. Let \( Z \in \mathbb{R}^{(E-1) \times K} \) be the matrix of target output states, where \( K \) is the number of output units.  

\NI To account for biases in the readout layer, a column of ones is appended to \( Y \), resulting in the matrix \( Y_{\text{bias}} \in \mathbb{R}^{(E-1) \times (N+1)} \):  
\[
Y_{\text{bias}} = \begin{bmatrix} Y & \mathbf{1}_{E-1} \end{bmatrix}
\]
where \( \mathbf{1}_{E-1} \in \mathbb{R}^{(E-1) \times 1} \) is a column vector of ones.  

\NI The weights and biases of the readout layer are computed by solving the following equation using the pseudoinverse of \( Y_{\text{bias}} \):  
\begin{equation}
W_{\text{bias}} = Y_{\text{bias}}^+ Z
\label{optWei}
\end{equation}
where \( Y_{\text{bias}}^+ \) is the Moore-Penrose pseudoinverse of \( Y_{\text{bias}} \), and \( W_{\text{bias}} \in \mathbb{R}^{(N+1) \times K} \) contains both the readout weights and the biases.  

\NI To compute the pseudoinverse, we first perform a singular value decomposition (SVD) of \( Y_{\text{bias}} \):  
\[
Y_{\text{bias}} = U S V^\top
\]
where \( U \in \mathbb{R}^{(E-1) \times (E-1)} \) is a unitary matrix, \( S \in \mathbb{R}^{(E-1) \times (N+1)} \) is a diagonal matrix containing the singular values, and \( V^\top \in \mathbb{R}^{(N+1) \times (N+1)} \) is the transpose of a unitary matrix.  

\NI The pseudoinverse of \( Y_{\text{bias}} \) is computed as:  
\[
Y_{\text{bias}}^+ = V S^+ U^\top
\]
where \( S^+ \in \mathbb{R}^{(N+1) \times (E-1)} \) is the pseudoinverse of the diagonal matrix \( S \). The pseudoinverse \( S^+ \) is obtained by taking the reciprocal of all non-zero singular values in \( S \) and leaving zeros unchanged.  

\NI Finally, after inserting \( Y_{\text{bias}}^+ \) into Eq.~\ref{optWei}, the optimal readout weights \( W \in \mathbb{R}^{K \times N} \) and biases \( b \in \mathbb{R}^K \) are extracted from the extended matrix \( W_{\text{bias}} \) as  
\[
W = \left( W_{\text{bias}}^\top \right)_{1:N,\,:}
\]
\[
b_w = \left( W_{\text{bias}}^\top \right)_{N+1,\,:}
\]
where the first \( N \) rows of \( W_{\text{bias}}^\top \) define the readout weights and the last row defines the biases.

\subsection{Fluctuation Measure}

\NI The neural fluctuation measure \( F \) quantifies the average temporal variability of reservoir activations. For each neuron \( n \), we compute the standard deviation \( \sigma_n \) of its activation time series \( y_n^{(t)} \). The global fluctuation is defined as the mean over the standard deviations of the individual neurons:
\[
F = \left\langle \sigma_n \right\rangle_n
\]

\NI Since $\tanh$-neurons produce outputs in \([-1, +1]\), the fluctuation \( F \) lies in \([0, 1]\). A value of \( F = 0 \) indicates a resting or fixed point state, while \( F = 1 \) corresponds to perfect two-state oscillation (e.g., alternating between $+1$ and $-1$).


\subsection{Covariance Measure}

\NI To assess temporal covariances, we compute the average product of the activation of neuron $m$ at time $t$ and neuron $n$ at time $t+1$:
\[
C_{mn} = \left\langle y_m^{(t)} \cdot y_n^{(t\!+\!1)} \right\rangle_t
\]

\NI Unlike the Pearson correlation coefficient, we deliberately avoid subtracting the mean or normalizing by the standard deviations. This ensures that the matrix elements $C_{mn}$ remain well-defined even when one or both signals are constant, as in a fixed point state.

\NI The global covariance measure is defined as the average over all neuron pairs, without differentiating between diagonal and off-diagonal elements:
\[
C = \left\langle C_{mn} \right\rangle_{mn}
\]

\NI Owing to the bounded output of the $\tanh$ neurons, the covariance values $C$ always lie within the range $[-1, +1]$.


\subsection{Nonlinearity Measure}

\NI The shape of the activation distribution \( p(y) \) reflects whether the reservoir operates in a linear or nonlinear regime. A central peak at \( y = 0 \) indicates a linear regime; two peaks near \( \pm1 \) indicate saturation and thus nonlinearity.

\NI We define a nonlinearity measure
\[
N = f_A - f_B + f_C
\]
based on the fractions of neural activations falling into the following intervals:
\[
\begin{array}{ll}
f_A & \in [-1,\;-0.5) \\
f_B & \in [-0.5,\;+0.5] \\
f_C & \in (+0.5,\;+1]
\end{array}
\]

\NI The resulting measure \( N \in [-1, +1] \) distinguishes three regimes:  
$N \approx -1$ for linear operation, $N \approx 0$ for intermediate or flat activation, and $N \approx +1$ for saturated, digital-like behavior.

\NI This intuitive yet robust definition proved most effective among several tested alternatives. It captures the essential qualitative transition in \( p(y) \) from unimodal (linear) to bimodal (nonlinear) distributions, as highlighted in earlier studies.


\subsection{Accuracy Measure}

\NI In the sequence generation task, we evaluate performance by comparing the actual output sequences \( Z_{\text{act}} \) of the readout layer with the corresponding target sequences \( Z_{\text{tar}} \), and compute the root-mean-square error \( E_{\text{RMS}} \). This error is normalized by the standard deviation \( \Delta Z_{\text{tar}} \) of the target sequences.

\NI To obtain an accuracy measure \( A \in [0, 1] \), we define:
\[
A = \frac{1}{1 + (E_{\text{RMS}} / \Delta Z_{\text{tar}})}
\]

\NI Note that \( A \approx 0.5 \) when the RMS error is comparable to the variability of the target data, and \( A = 1 \) when the output \( Z_{\text{act}} \) matches the target \( Z_{\text{tar}} \) exactly.

\NI In classification tasks, the accuracy $A$ is simply defined as the fraction of correctly predicted class labels.

\newpage
\section{Results}

\subsection{Naming Conventions and Essential Methods}

\NI In the following section, we briefly introduce the naming conventions as well as the tools for control, visualization, and analysis that are used throughout the paper. Further details are provided in the Methods section.

\paragraph{Framework:} We consider a Reservoir Computer (RC) consisting of an input matrix, a Recurrent Neural Network (RNN) serving as the reservoir, and a readout layer trained using the pseudoinverse method. To make multiple runs feasible for many parameter combinations, we restrict ourselves to reservoirs with only 50 neurons. This choice also allows us to draw on experience from our previous publications, where the same reservoir size was mostly used.

\paragraph{Statistical Control Parameters:} The elements $w_{ij}$ of the weight matrix describing the recurrent connections among the reservoir neurons are random, but obey three {\bf statistical control parameters}. The {\bf density} $d\in\left[0,1\right]$ denotes the fraction of non-zero elements. In this work we restrict ourselves to fully connected networks with $d\is1$. The {\bf coupling strength} $w$ is defined as the standard deviation of the normally distributed weight matrix elements (in a balanced system). The {\bf balance} $b\in\left[-1,+1\right]$ characterizes the relative predominance of excitatory and inhibitory connections: $b\is-1$ corresponds to exclusively negative weights, $b\is0$ to a balanced system, and $b\is+1$ to exclusively positive weights. A reservoir described by these three parameters and without further modification is called {\bf homogeneous}, because all matrix elements follow the same statistical rule.

\paragraph{Weight Matrix Structuring:} Without changing its global weight distribution $p(w)$, a homogeneous weight matrix can be structured by redistributing its elements across the matrix in various semi-regular ways. For example, a geometrically defined subset of the matrix, such as a small random fraction $f_1$ of all matrix rows, may preferentially receive more extreme elements from the distribution $p(w)$, such as those with the weakest or strongest magnitudes, or those with the most negative or most positive values. Instead of selecting random rows as the $f_1$ subset, one can also use columns or quadratic blocks of size $S$ along the matrix diagonal. We refer to this distribution-conserving method of patterning the weight matrix as {\bf 'Permutative Structuring'}. Some examples of homogeneous and structured matrices are shown in Fig.\ref{fig2}.

\paragraph{Neural Activity Vizualization:} The dynamical processes in the reservoir can be most directly visualized by plotting the {\bf individual activations} of all neurons over successive time steps (compare Fig.\ref{fig1}(b)). Since the neurons use tanh activation functions, their continuous outputs are always in the range from $-1$ to $+1$. However, when the system develops strong spontaneous dynamics and runaway excitations drive the neurons deep into saturation, it becomes impossible to see the very small input-related perturbations that may still be superimposed on the spontaneous activity and that could nevertheless be exploited by the readout layer.

\paragraph{Principal Component Analysis:} For this reason, we also perform a Principal Component Analysis (PCA) on the entire temporal evolution of the reservoir states and then plot the dominant {\bf PCA components} as a function of time (compare Fig.\ref{fig3}). The different PCA components can vary in magnitude by many orders of magnitude. Since the readout layer is insensitive to these differences, we normalize all components individually within the displayed time window to the range from $-1$ to $+1$. This normalization makes even weak patterns visible that are clearly related to the input signals (rather than to the spontaneous dynamics), because they change in synchrony with the pulses of the input episodes.

\paragraph{Dynamical Measures:} For a more quantitative analysis, we compute four aggregating dynamical measures from the entire, multi-episode temporal evolution of the reservoir states: The {\bf fluctuation} $F\in\left[0,1\right]$ quantifies the global level of change in the neural activations, averaged both over the neural population and over time. Two {\bf covariances $C_0$ and $C_1$}, both ranging from $-1$ to $+1$, measure the (non-normalized) degree of correlation between neurons, averaged over all neuron pairs. Here, $C_0$ denotes the instantaneous correlation and $C_1$ the correlation at lag time one. Finally, the {\bf nonlinearity} $N$, ranging from $-1$ to $+1$, quantifies whether the neurons operate predominantly in the linear or in the nonlinear, saturated part of their sigmoidal activation function. Values close to $-1$ indicate linear behavior, whereas values close to $+1$ indicate saturated behavior. Together, these four measures allow us to classify the reservoir's dynamical state into four broad regimes, which we refer to as 'oscillatory', 'chaotic', 'calm', and 'fixed point'. It is particularly instructive to plot these measures as functions of some control parameter that is swept through its range and to observe how the RNN is switching between dynamical regimes accordingly (compare Fig.\ref{fig4}).

\paragraph{Task and Accuracy:} We are primarily interested in the question of how structural properties of the weight matrix (or regulating control mechanisms) influence the dynamical regime of the RNN and which of these regimes are suitable for task-related information processing. The quality of information processing, in turn, is most directly assessed through the accuracy achieved in specific tasks. This poses a challenge, however, because it is nearly impossible to take into account all possible types of tasks with their different information-processing requirements. In this work, we therefore consider only a single pseudo-task, called 'Sequence Generation'. The RC receives sequences of input vectors belonging to distinct classes and has to respond with predefined output sequences that are specific to the respective input class. The components of both input and output vectors are continuous numbers between $-1$ and $+1$. Although this task may not be particularly useful for real-world applications, it provides a clear test of whether the reservoir dynamics are mainly determined by the most recent input sequence, by remnants of earlier input episodes, or by spontaneous dynamics with the potential for runaway excitation. To assess performance in the task, we first compute the root-mean-squared error between the target and the actual output vectors and then convert this error into an {\bf accuracy} $A$. This accuracy is approximately $0.5$ at chance level and approaches $1$ when performance is near perfect (compare black lines in Fig.\ref{fig4}).

\subsection{Problem of Runaway Excitation}

\paragraph{Weakly Coupled Reservoirs:} It is instructive to first consider a reservoir with relatively weak recurrent connections, such as in the case of a coupling strength $w\is0.1$ (compare Fig.\ref{fig4}(a)). In this situation, the dynamical regime of the RNN depends on the balance parameter $b$. 

\NI When $b$ is close to zero, corresponding to a balanced weight matrix with equal amounts of negative and positive entries, the weakly coupled RNN operates in a calm regime. This regime is characterized by vanishing covariances $C_0,C_1$, small fluctuations $F$, and a nonlinearity parameter $N$ close to $-1$, indicating that neurons operate mainly in the linear regime. The dynamics are primarily determined by the most recent input, there is no spontaneous activity, and consequently the accuracy $A$ reaches nearly perfect values.

\NI However, when the balance is tuned too far into the positive range, so that most connections become excitatory, herd effects emerge and positive feedback drives the system into a global fixed-point attractor. In this regime, fluctuations $F$ nearly vanish, both covariances approach $+1$, and all neurons are driven into saturation, indicated by a nonlinearity parameter $N$ approaching $+1$. In general, such saturated, locked-in neurons are not well suited for information processing, because the trajectory of the reservoir state should be able to reflect changes in the input signals. Nevertheless, in the weakly coupled RNN, tiny input-related differences in neural activations remain present even in the saturated neurons. The readout layer can still exploit these differences and, remarkably, the accuracy $A$ remains close to perfect even in the fixed-point regime.

\NI Another herd effect occurs when the balance is tuned too far into the negative range, so that most connections become inhibitory. In this case, negative feedback drives the system into global oscillations with period two, in which all neurons alternately switch between negatively and positively saturated states. Again, this situation would normally be detrimental for information processing, but due to the weak coupling the accuracy remains close to perfect even in this oscillatory regime. 

\paragraph{Strongly Coupled Reservoirs:} The ability to perform input-related information processing collapses almost completely when the coupling strength is increased to $w\is1$ (compare Fig.\ref{fig4}(b)).

\begin{figure}[ht!]
\centering
\includegraphics[width=0.7\linewidth]{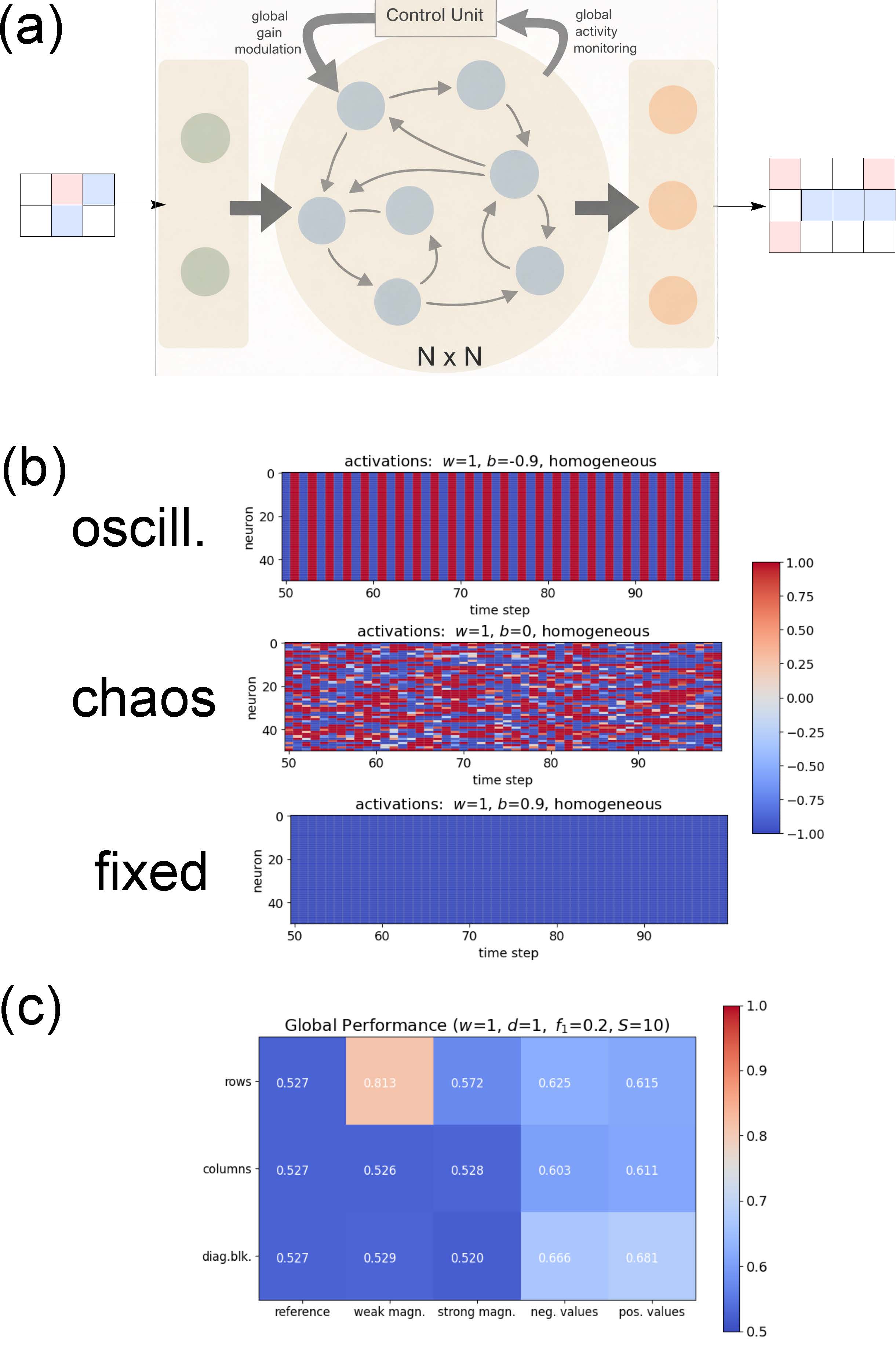}
\caption{{\bf Overview}.
\textbullet$\;${\bf(a)} Reservoir computer consisting of a recurrent neural network (RNN with $N$ neurons, central circle with blue dots), an input layer (left box with green dots), and a readout layer (right box with orange dots). A control unit for dynamic gain regulation can optionally be placed on top of the RNN. The colored grids symbolize sequences of input and output vectors.
\textbullet$\;${\bf(b)} Three direct plots of neural activations (color coded) over time in a densely connected RNN of 50 neurons with tanh activation functions. Because the coupling strength is large ($w\is1$), the system develops global oscillations for strongly negative balance parameters such as $b\is-0.9$, chaotic fluctuations for highly balanced systems such as $b\is0$, and global fixed points for strongly positive balance parameters such as $b\is+0.9$. All these attractor regimes are detrimental to task-related information processing.
\textbullet$\;${\bf(c)} The Global Performance (GP, color coded) measure, defined as the accuracy averaged over nine bias values spanning the full range from $-1$ to $+1$, evaluated for $3\times5$ different types of permutative matrix structuring that all preserve the global weight distribution (compare main text and \ref{fig2}). The best GP is obtained when 20 percent of the rows in the weight matrix (representing neural inputs) have a weaker-than-average magnitude.
}
\label{fig1}
\end{figure}

\NI Even when the balance $b$ is around zero, the RNN is no longer calm. Instead, the strong recurrent bipolar connections lead to spontaneous chaotic fluctuations (compare Fig.\ref{fig1}(b), middle plot), which are not related to the momentary external input. In this chaotic regime, both covariances $C_0$ and $C_1$ are close to zero, whereas the fluctuation $F$ and the nonlinearity $N$ approach the upper limit $+1$. As a consequence, the accuracy $A$ drops to chance level $0.5$.

\NI In the vicinity of $b\is-1$ and $b\is+1$, we again observe saturating oscillatory or fixed-point behavior (compare Fig.\ref{fig1}(b), upper and lower plot), respectively. In contrast to the weakly coupled RNN, however, the accuracy now remains close to chance level also in these highly unbalanced systems.

\NI This leaves the two regions between chaos and oscillatory behavior, as well as between chaotic and fixed-point behavior, as the only transitional regimes in which the accuracy $A$ can rise to moderately satisfactory levels. This confirms the well-known hypothesis that the 'edge of chaos' constitutes a sweet spot for information processing. In practical applications, however, this would require careful tuning of the balance parameter in order to achieve optimal performance.

\NI Note that in the vicinity of $b\is-1$ and $b\is+1$, the four dynamical measures $F$, $C_0$, $C_1$, and $N$ have very similar values in the weakly and strongly coupled RNN (compare Fig.\ref{fig4}(a,b)). Direct plots of the neural activations also appear identical (as in the upper and lower plots of Fig.\ref{fig1}(b)) in strongly unbalanced systems, both in the weakly and strongly coupled case. Yet the accuracy $A$ reaches perfection in one case and remains at chance level in the other. This clearly indicates that our dynamical measures and direct activation plots are not sensitive enough to reveal the presence or absence of tiny input-related perturbations superimposed on the global herd behavior. For this reason, we use in the following sections Principal Component Analysis with visual magnitude enhancement to reveal these subtle effects.

\paragraph{Statement of the Problem:} We are now in a position to formulate the problem and research goal addressed in this work: In strongly coupled RNNs, herd effects prevent information processing both in strongly unbalanced systems and in the perfectly balanced system. How can this limitation be overcome without fine-tuning the balance parameter to the narrow sweet spot at the edges of chaos?

\subsection{Permutative Structuring of the Weight Matrix}

\paragraph{Previous Work:} In a previous publication (\cite{metzner2025organizational}) we demonstrated that Dale's principle (requiring that every output column in the weight matrix should be either strictly positive or strictly negative) has a small performance-enhancing effect, whereas the reciprocal symmetry of Hopfield networks (requiring that $w_{ij}$ should be similar to $w_{ji}$) generally tends to degrade performance.

\paragraph{Leptokurtic Weight Distributions:} In the same paper we also observed a modest performance gain when introducing quadratic patches into the weight matrix, some of which contained larger-than-average magnitudes and others smaller-than-average magnitudes. Although the overall standard deviation was strictly conserved in that investigation, the patch-like patterning made the overall distribution more leptokurtic, thereby increasing the fraction of near-zero weights. As it turned out after publication, this effective reduction of the connection density $d$ was mainly responsible for the observed performance enhancement. 

\paragraph{Improved Patterning Method:} For this reason, we have now developed a new method of matrix patterning that strictly conserves the entire weight distribution, because it is based on a controlled permutation of the existing weights across the matrix (see Methods section for details).

\paragraph{Comparing Types of Structuring:} We apply four different kinds of modifications (weaker-than-average magnitudes, stronger magnitudes, more negative values, and more positive values) to a small subset of the matrix elements, arranged in three different geometric patterns (rows, columns, and diagonal blocks), thus creating twelve different combinations of permutative structuring. For each combination, we compute a global performance measure GP, defined as the accuracy averaged over nine equally spaced balance values $b$ in the range from $-1$ to $+1$. The baseline is the GP obtained for the non-structured, homogeneous matrix, which is 0.527. The results are summarized in Fig.\ref{fig1}(c).

\paragraph{Weak Rows:} We find the strongest performance enhancement (raising the GP from 0.527 to 0.813) when weaker-than-average weight magnitudes are assigned to 20 percent of randomly selected rows in the weight matrix. Consequently, the remaining 80 percent of rows contain larger-than-average magnitudes. Since the matrix rows represent the input weights of the neurons, this best-performing RNN contains a subpopulation of 20 percent of neurons that receive particularly weak inputs from the rest of the network. In situations where the network suffers from herd effects that generate runaway excitations and lead to performance-degrading saturation, this subpopulation can remain partially decoupled from the global network dynamics and still provide useful input-related information to the readout layer.

\paragraph{Unipolar blocks:} The next best performance enhancement (raising the GP from 0.527 to 0.666 or 0.681, respectively) is observed when all defined diagonal blocks of linear size 10 receive preferentially negative or preferentially positive values. A positive block, for example, corresponds to a kind of module with predominantly excitatory internal connections that is embedded in a larger surrounding network with the opposite polarity of connections.

\paragraph{Overview and Visualization:} Of the 12 investigated combinations of matrix structuring, 8 lead to an enhancement of the GP. The top three combinations are shown for different bias values $b$ in Fig.\ref{fig2}(a). The four combinations that do not increase performance are shown in Fig.\ref{fig2}(b).

\subsection{Effect of Weak Rows}

\NI We next investigate the effect that a small subset of neurons with weak input weights has on the dynamics of the reservoir and how this influences its information-processing capability.

\begin{figure}[ht!]
\centering
\includegraphics[width=0.95\linewidth]{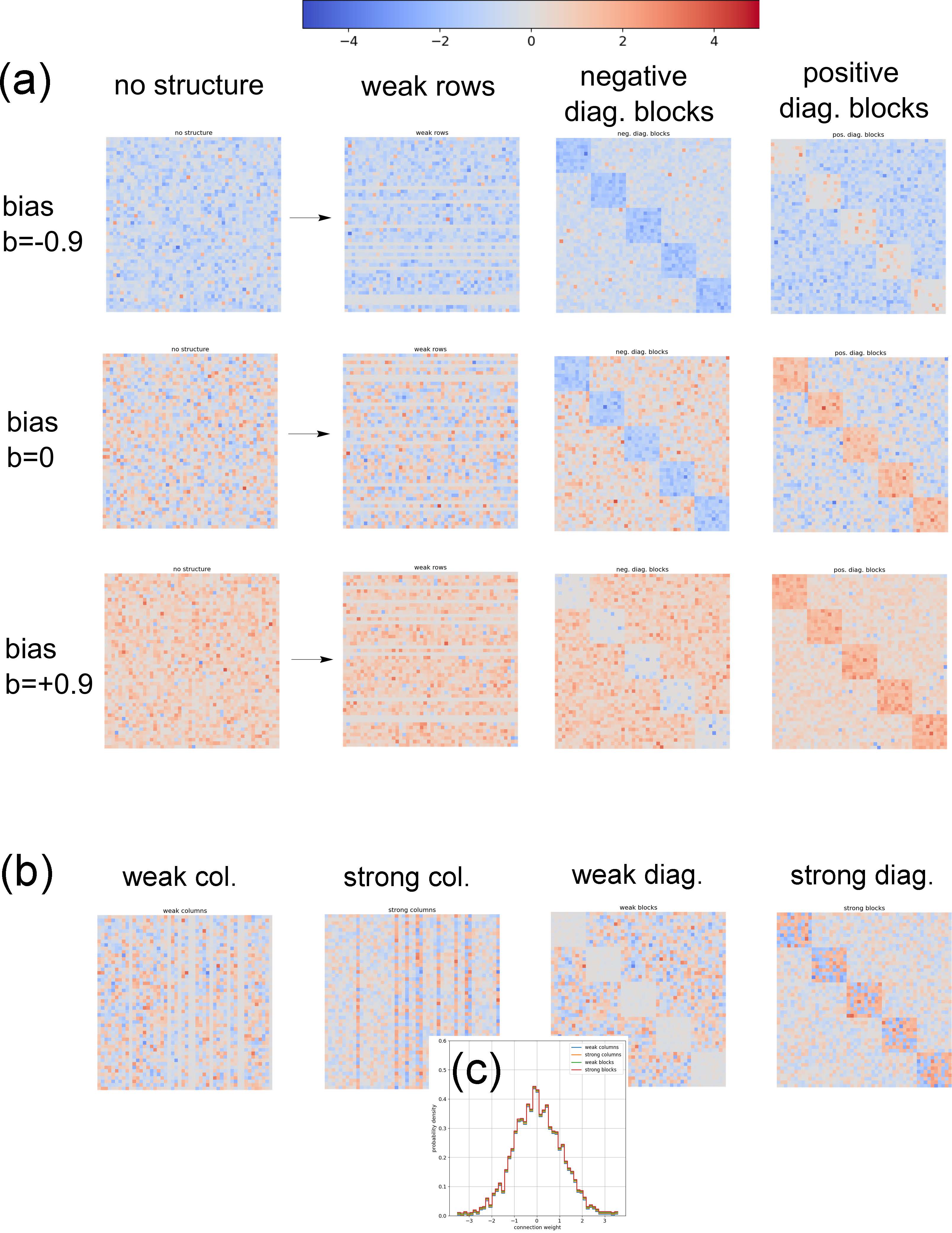}
\caption{{\bf Structuring of the weight matrix}.
\textbullet$\;${\bf(a)} Three examples of GP-enhancing types of permutative matrix structuring, shown for three different balance values $b$. In each case, the full reservoir connection matrix is displayed with color-coded weights.
\textbullet$\;${\bf(b)} Four examples of permutative matrix structuring that do not lead to an improvement in global performance.
\textbullet$\;${\bf(c)} Probability distributions of connection weights for four different types of permutative matrix structuring. The distributions remain unchanged. The four curves have been slightly shifted vertically to improve visibility.
}
\label{fig2}
\end{figure}

\paragraph{Dynamics as a function of bias:} For this purpose, we again perform a scan of the bias parameter $b$ over its entire range and compute the four dynamical measures as functions of $b$ (compare Fig.\ref{fig2}(c)). When compared to the corresponding homogeneous system (Fig.\ref{fig2}(b)), the quantities $F$, $C_0$, $C_1$, and $N$ exhibit very similar behavior. Only in the chaotic regime around $b\is0$ do the weak rows lead to a significant reduction of the fluctuation $F$ and the nonlinearity $N$, which is reflected in an increase of the accuracy $A$ to slightly above chance level. Remarkably, in the strongly unbalanced regimes close to $b\is-1$ and $b\is+1$, the weak rows allow the accuracy $A$ to rise to nearly the perfect level of 1, even though no difference is visible in the four dynamical measures. This again demonstrates that these measures are well suited for assessing the dynamical regime of the reservoir, but not for evaluating its information-processing capabilities.

\paragraph{PCA-Analysis of the Weak Row Effect:} To gain further insight into the effect of weak rows, we next compare the time dependence of the reservoir's lowest PCA components with those of the corresponding homogeneous system (Fig.\ref{fig2}, left and middle columns of plots). Here, we are mainly interested in the balance regions that are unsuitable for information processing in the homogeneous system. We therefore restrict this PCA analysis to five representative balance values $b\in\left\{-1,-0.9,0,+0.9,+1\right\}$.

\paragraph{PCA-Components in the Homogeneous RNN:} In the homogeneous RNN at $b\is-1$ and $b\is-0.9$, we observe the typical period-two global oscillation in all of the lowest eight PCA components (Fig.\ref{fig2}, left column, upper two plots). Recall that the time series of each component is separately rescaled to the range from $-1$ to $+1$ for better visibility. Nevertheless, the plot reveals that the values within each time series flip in a strictly regular way between only two binary states and therefore do not provide the input-dependent differences required by the readout layer.

\NI In the balanced case at $b\is0$, all components exhibit chaotic behavior (left column, middle plot). These fluctuations are likewise not sufficiently determined by the input signals.

\NI Finally, at $b\is+0.9$ and $b\is+1$, we have seen before that the RNN resides in a global fixed point, in which all neuron activations remain at the same constant, highly saturated value. Since this constant activation vector --- corresponding to the mean or `center point' of the high-dimensional `cloud' of momentary activation vectors --- is subtracted in the PCA analysis, all components take the value zero (left column, lower two plots). Clearly, this again provides no useful information for the readout layer.

\paragraph{PCA-Components in the RNN with Weak Rows:} The introduction of a few weak rows in the weight matrix does not prevent the PCA components from appearing chaotic in the perfectly balanced system at $b\is0$ (Fig.\ref{fig2}, middle column, center plot). Nevertheless, recall that even in this chaotic regime the accuracy is now slightly above chance level, meaning that some input-related information must be hidden within these irregular fluctuation of the PCA components.

\NI The situation changes much more dramatically for the other four considered bias values (middle column, upper and lower plots). The momentary values of the PCA components are no longer restricted to only three possible values ($-1$, $0$, and $+1$, after rescaling). At least some of the higher components can now choose from a broader and more informative domain of values. Moreover, in some components one can observe a quasi-periodic, regular pulsing with a period of three, corresponding to the length of each input episode. Closer inspection reveals two broad categories of pulses, reflecting the two classes of input vectors in the Sequence Generation task. Even within pulses of the same category, however, the values differ slightly. This richer, input-related information content in the PCA components enables the readout layer to produce the target output sequences with high accuracy, even when the system is globally in an oscillatory or fixed-point regime.

\subsection{Effect of Automatic Gain Control (AGC)}

\paragraph{Motivation of AGC:} As we have seen, suitable patterning of the weight matrix, without altering the distribution of its elements, can make a strongly coupled reservoir capable of information processing even in cases of large imbalance between excitatory and inhibitory connections. In the chaotic regime, the patterning also slightly increases the accuracy to values above chance level. Nevertheless, in a practical setting one would still attempt to keep the balance away from the chaotic regime in order to maximize information-processing performance. We therefore seek additional extensions of the reservoir computer that make its performance less dependent on the detailed statistical and structural properties of the weight matrix.

\paragraph{Inspiration by the Brain:} In particular, we consider an external control unit that continuously monitors the dynamical state of the reservoir and, if necessary, sends back regulatory feedback signals in order to keep the network dynamics close to a desired target state (Fig.\ref{fig1}(a), top). Inspired by biological neural systems, where the overall excitability and effective coupling strength of neurons are modulated by neuromodulators, hormones, or glial interactions, we allow only global (not neuron-specific) signals to be exchanged between the control unit and the reservoir of neurons.

\paragraph{Implementation of AGC:} As described in more detail in the Methods section, we aim to regulate the global activity of the reservoir, quantified as a running time average of the momentary root-mean-squared activation across all neurons, to a predefined target value. To allow the control unit to influence this global activity, we do not modify the entries of the weight matrix individually, but instead modulate a single global prefactor of that matrix according to the deviation between the actual and the target activation. This mechanism can also be interpreted as a dynamical regulation of the effective coupling strength $w$.   

\begin{figure}[ht!]
\centering
\includegraphics[width=1\linewidth]{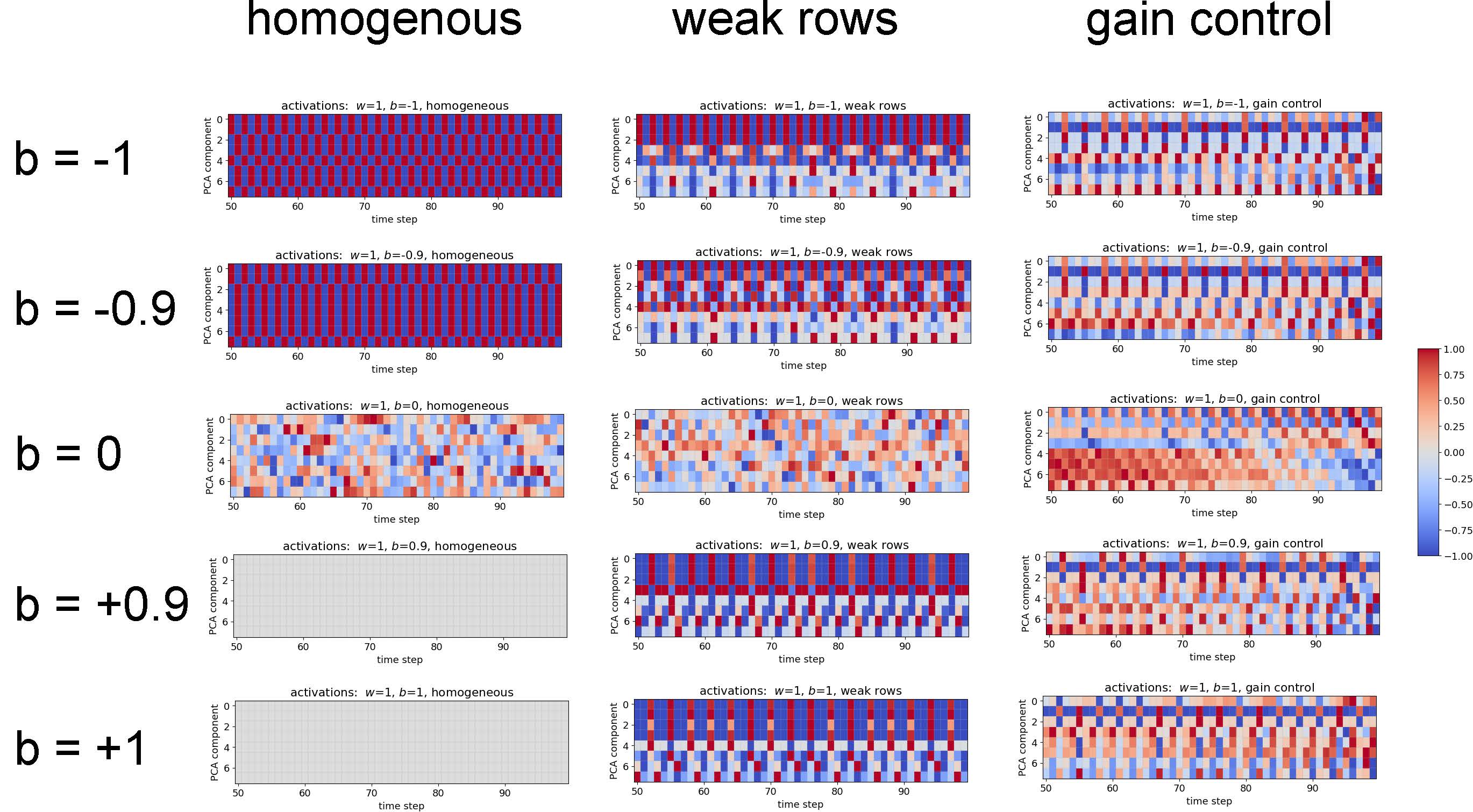}
\caption{{\bf Principal Component Analysis (PCA) of reservoir activations.}. Each of the $5\times3$ plots represents the lowest eight principal components of the reservoir activation states as a function of time. The magnitudes of these components differ greatly. Since the readout layer is insensitive to absolute magnitude, we normalize the values within the observation window to the full range from $-1$ to $+1$ to improve visibility.
\textbullet$\;$ The first column of plots shows the homogeneous (non-structured) system. In the range of strongly negative bias ($b\is-1$ and $b\is-0.9$), all PCA components reflect the reservoir's global oscillations with a period of two. In the balanced case ($b\is0$), the components fluctuate chaotically. In the range of strongly positive bias ($b\is+0.9$ and $b\is+1$), where the reservoir resides in a global fixed-point attractor, all PCA components are zero because the mean is subtracted in the PCA analysis.
\textbullet$\;$ The second column of plots shows the structured system with some rows of reduced magnitude in the weight matrix. In the balanced case ($b\is0$), we again observe chaotic behavior. However, in all strongly unbalanced cases ($b\is-1$, $b\is-0.9$, $b\is+0.9$ and $b\is+1$), the PCA components exhibit much richer dynamics. Some display a quasi-periodic pattern with a period of three, corresponding to the length of each input episode. These components therefore encode input-related information that can be exploited by the readout layer.
\textbullet$\;$ The third column of plots shows the system equipped with a unit for automatic gain control. Here, the information encoded in the PCA components is even more diverse and more strongly aligned with the pulses of input.
}
\label{fig3}
\end{figure}

\begin{figure}[ht!]
\centering
\includegraphics[width=0.7\linewidth]{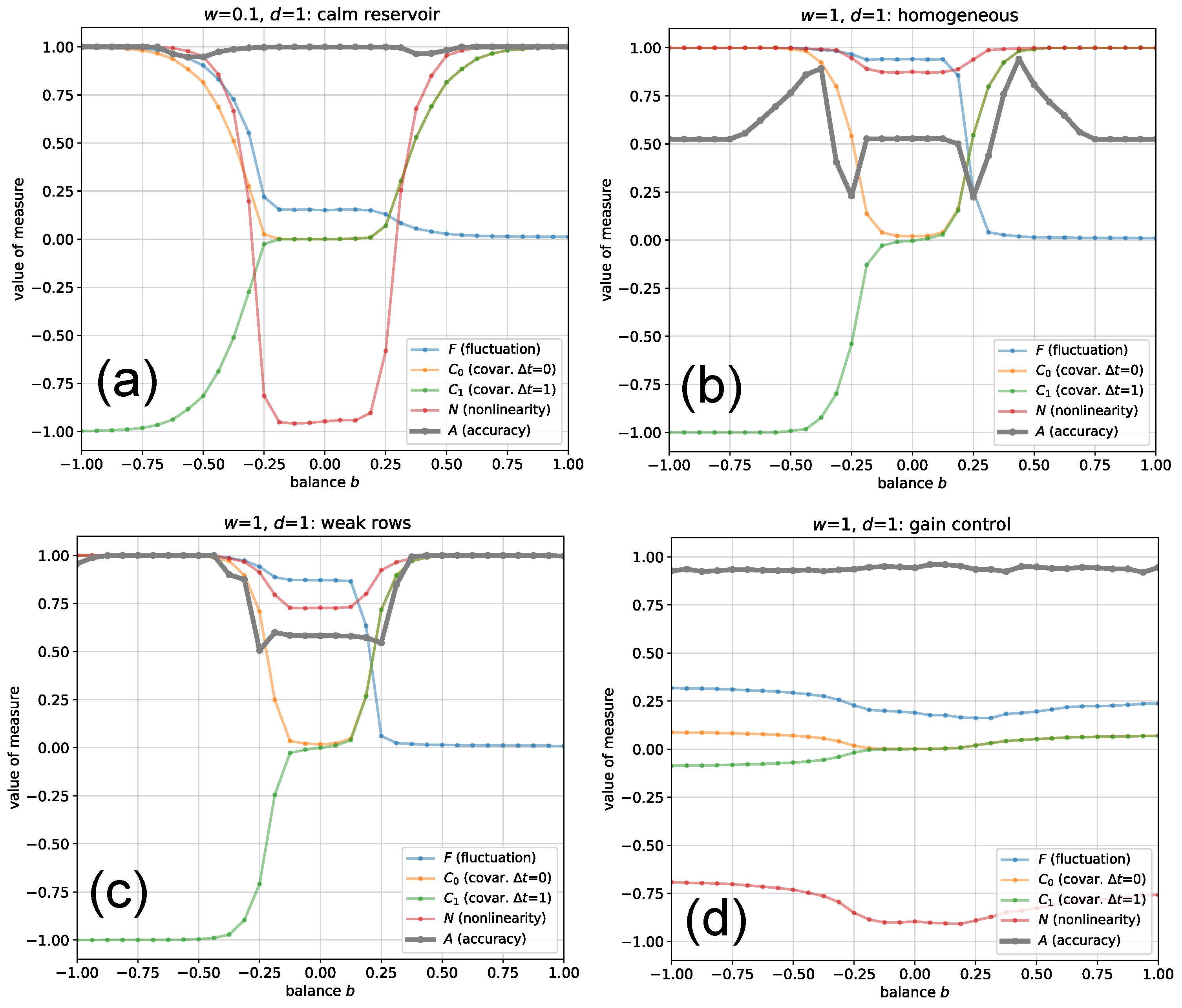}
\caption{{\bf Scan of dynamical properties and accuracy over the excitatory/inhibitory balance parameter $b$.}
\textbullet$\;${\bf(a)} Non-structured system with the coupling strength reduced to $w\is0.1$. Even in this weakly coupled reservoir, three distinct dynamical regimes appear. In the globally oscillating regime at strongly negative $b$, the fluctuation $F$ (blue), the nonlinearity $N$ (red), and the instantaneous covariance $C_0$ (orange) approach the value $+1$, whereas $C_1$ (green), the covariance at lag time one, approaches $-1$. In the calm regime around $b\is0$, fluctuations $F$ are small, the nonlinearity parameter $N$ is close to $-1$ (indicating that neurons operate in the linear regime of the activation function), and both covariances are close to zero. In this calm regime, the reservoir dynamics are mainly determined by the external input signals. In the global fixed-point regime at strongly positive $b$, the nonlinearity $N$ and both covariances $C_0$ and $C_1$ approach $+1$, whereas the fluctuation $F$ becomes zero. Although the neurons are saturated in the oscillatory and fixed-point regimes, the accuracy $A$ remains close to $1$ for all balances $b$, due to the weak coupling strength $w\is0.1$. In weakly coupled reservoirs, information-processing activity can "ride on top" of these large activations, as we have shown previously.
\textbullet$\;${\bf(b)} Non-structured system with the coupling strength increased to $w\is1$. Within the oscillatory and fixed-point regimes, the dynamical properties $F$, $C_0$, $C_1$, and $N$ behave qualitatively as in the weakly coupled reservoir. However, in the balanced system around $b\is0$, the calm system state is now replaced by high-amplitude chaotic spontaneous fluctuations. This leads to vanishing covariances $C_0\is C_1\is0$ and values close to one for the fluctuation $F$ and the nonlinearity $N$ (indicating operation in the saturation regime of the activation function). The accuracy $A$ is now close to chance level $A\is0.5$ in the oscillatory, chaotic, and fixed-point regimes. Only at the two "edges of chaos" does the accuracy rise to levels of good performance.
\textbullet$\;${\bf(c)} Structured system with 20 percent weak rows. Here, the dynamical quantities appear qualitatively similar to case (b), but in the chaotic regime the fluctuation $F$ and the nonlinearity $N$ are significantly reduced. Most importantly, however, the accuracy $A$ is now close to one for all balances $b$ except in the chaotic regime, where it remains only slightly above chance level.
\textbullet$\;${\bf(d)} System with automatic gain control. Here, the nonlinearity parameter $N$ is relatively close to $-1$, indicating predominantly linear operation of the neurons. Fluctuations $F$ are small and mainly input-driven. The covariances are very small. Together, the dynamical quantities resemble those of the weakly coupled reservoir in panel (a) within the calm regime around $b\is0$, yet this calm state now extends over the full range of balance values between $-1$ and $+1$. The accuracy is also close to perfect across this full balance range.
}
\label{fig4}
\end{figure}

\paragraph{Dynamics and Performance with AGC:} We apply this method of AGC to the strongly coupled system with $w\is1$ and again scan the balance parameter $b$ across its entire domain (Fig.\ref{fig4}(d)). We find that all four dynamical parameters $F$, $C_0$, $C_1$, and $N$ now exhibit behavior that qualitatively resembles that of the weakly coupled ($w\is0.1$) reservoir in the calm ($b\is0$) regime, but now across the entire range of balance values $b$. Fluctuations $F$ remain close to 0.25 (the chosen target value of the reservoir activation) and the nonlinearity $N$ remains close to $-1$, indicating that neurons operate mainly in a linear or only weakly nonlinear region of the activation function. Both covariances $C_0$ and $C_1$ remain small in magnitude, indicating that both oscillatory and fixed-point behavior are strongly suppressed by the continuous dynamical regulation of reservoir activity. Most importantly, the accuracy $A$ remains very high for all values of $b$, although slightly below perfect performance. A visualization of the time-dependent PCA components (Fig.\ref{fig3}, right column) reveals that the reservoir now produces a set of diverse states that appear even richer in information than those of the system with weak rows (Fig.\ref{fig3}, middle column). Again, many of the PCA components show the pulsing behavior with period three, indicating the representation of input-related content. This remains true even in the balanced system at ($b\is0$), because due the AGC there is no chaotic regime any longer. 

\section{Conclusions}

\subsection{Summary}

\NI In the context of reservoir computing (RC), we address in this paper the problem of runaway excitations in recurrent neural networks (RNNs). These excitations arise from herd effects, particularly in dense neural networks with unbalanced excitatory and inhibitory connections. As a result, the network is driven into global oscillatory or fixed-point attractors in which all neurons are pushed into the flat, saturating region of their sigmoidal activation functions—an operating regime that is largely insensitive to external input signals.

\NI As long as the overall coupling strength of the neural network remains sufficiently small, tiny input-related perturbations can still “ride on top” of these high-amplitude global attractors, allowing the readout layer to exploit these perturbations for the task at hand. However, in strongly coupled networks the neurons are driven so deeply into saturation that processing of input-related information becomes impossible, and the performance of the reservoir computer degrades drastically.

\NI A related problem also emerges in balanced systems when the coupling strength becomes too large. In weakly coupled balanced RNNs, the dynamics is calm, quasi-linear, and primarily driven by external input, which is ideal for many information-processing tasks. In contrast, strongly coupled balanced RNNs transition from this calm regime into a chaotic regime that is also detrimental to input-related information processing.

\NI Consequently, in strongly coupled reservoir computers the performance (for example in a task such as sequence generation), when plotted as a function of the balance parameter, remains close to chance level across most of the parameter range. Only two narrow peaks appear at the so-called “edges of chaos,” making practical operation difficult because it requires careful fine-tuning of the network parameters.

\NI As a first countermeasure to these runaway excitations, we explored subtle ways of structuring the weight matrix without altering the overall distribution of its elements. In particular, we introduced a small random fraction of rows whose elements have smaller-than-average magnitudes. This modification corresponds to a subpopulation of neurons that receive only relatively weak input signals from the rest of the network, and it proved to be a particularly effective stabilization mechanism. As a result of this “minimally invasive,” permutative modification of connectivity, the accuracy becomes excellent even in the largely unbalanced regimes characterized by oscillatory and fixed-point attractors, whereas it shows little improvement in the chaotic regime.

\NI Besides the introduction of “weak rows,” the performance of strongly coupled systems could also be slightly enhanced by introducing other types of patterns into the weight matrix, such as preferentially negative or positive diagonal blocks. By contrast, we also demonstrated that several other possible forms of patterning had no performance-enhancing effect.

\NI Our ultimate goal, however, was to develop a countermeasure against runaway excitations that enables the RNN to perform meaningful information processing regardless of the excitatory/inhibitory balance. This was achieved by augmenting the reservoir computer with an additional unit that implements an Automatic Gain Control (AGC). Although this control unit receives only a global activity signal from the reservoir and sends back only a global modulatory signal into it, the resulting dynamic regulation keeps the RNN in a calm, input-receptive state, irrespective of the balance parameter. Accordingly, the accuracy remains at a very good level across the entire range of balances.

\subsection{Outlook}

\NI In future work, it would be interesting to test structural and dynamical countermeasures against runaway excitations across a broader range of tasks with different demands on information processing in the reservoir. Another open question concerns which kinds of tasks actually require strong coupling in the first place. In our previous publications, we have already identified several tasks that require only mildly nonlinear dynamics and would therefore also work in weakly coupled RNNs. On the other hand, there are tasks such as the prediction of binary systems that may work best in strongly coupled RNNs, where neurons operate in the saturated part of their activation functions and behave more like binary logic gates.

\NI This paper has focused on situations in which runaway excitations arise because the statistical control parameters of the weight matrix—such as the density $d$, the coupling strength $w$, and the balance $b$—are not optimally chosen. A practically even more relevant scenario would involve an RNN whose control parameters are optimized for processing typical inputs, but which can occasionally be pushed into a computationally unfavorable attractor by rare “bad” inputs. For example, the RNN might be tuned close to one of the edges of chaos and normally remain in a healthy dynamical regime, but transition into chaos when exposed to occasional inputs with very large amplitudes. In such cases, it would be particularly interesting to examine whether the AGC mechanism can restore the dynamics to the desired regime within a sufficiently short time interval. 

\subsection{Discussion}

\NI The results of the present study connect to several lines of research in reservoir computing, nonlinear dynamical systems, and computational neuroscience. In particular, they highlight how structural heterogeneity and dynamical regulation can influence the operating regime of recurrent neural networks and thereby affect their ability to perform input-related information processing.

\NI A central concept in reservoir computing is the so-called echo state property, which requires that the internal state of the network should eventually become determined primarily by the recent input history rather than by arbitrary initial conditions \cite{jaeger2001echo}. In practice, this property is often ensured by constraining the spectral radius of the recurrent weight matrix or by otherwise limiting the effective coupling strength within the reservoir. Subsequent work has further clarified the mathematical conditions under which this property holds and how it relates to the stability of the underlying dynamical system \cite{manjunath2013echo,yildiz2012revisiting}. Our results suggest an alternative perspective: instead of enforcing a fixed structural constraint on the network parameters, it may be possible to achieve a similar stabilizing effect through dynamical regulation. The automatic gain control mechanism implemented here continuously adjusts the effective coupling strength of the network in response to its momentary activity level. In this way, the system is actively maintained in a dynamical regime that remains sensitive to external input, even when the underlying connection statistics would normally push the network into chaotic or saturating attractor states.

\NI This idea of regulating network dynamics through feedback mechanisms also resonates with observations from biological neural systems. Neural circuits in the brain are not static networks with fixed operating parameters but are subject to multiple forms of homeostatic regulation that maintain their activity within a functional range. Examples include synaptic scaling and other forms of activity-dependent plasticity that adjust the overall excitability of neural populations \cite{turrigiano2012homeostatic}. Inhibitory plasticity has also been proposed as a mechanism that dynamically stabilizes excitation–inhibition balance in cortical circuits \cite{vogels2011inhibitory}. More generally, neuronal circuits exhibit a variety of regulatory processes that compensate for parameter variability and maintain stable functional regimes \cite{marder2006variability}. Although the gain control mechanism explored in this work is intentionally simple and operates through a single global control signal, it illustrates how even coarse-grained feedback can stabilize a large recurrent network and extend the regime in which useful computation is possible. In this sense, the mechanism can be interpreted as a minimal computational analogue of biological homeostatic regulation.

\NI The structural countermeasure investigated in this work—the introduction of a small subset of neurons receiving weaker-than-average inputs from the rest of the reservoir—offers a complementary perspective. In strongly coupled networks, runaway excitation typically arises from collective feedback loops that synchronize the activity of large parts of the network. The presence of neurons with weaker incoming connections effectively creates a subpopulation that remains partially decoupled from these global herd effects. Even when the majority of neurons are driven into oscillatory or fixed-point behavior, this weakly coupled subset can still respond to input-related perturbations and thereby preserve useful information in the reservoir state. The results therefore suggest that small degrees of structural heterogeneity can substantially enhance the robustness of reservoir dynamics.

\NI More generally, this observation relates to a broader theme in the study of complex networks: heterogeneous connectivity patterns can prevent the emergence of overly synchronized global dynamics and thereby promote richer dynamical behavior. In biological neural networks, connectivity is known to exhibit pronounced heterogeneity and modular organization \cite{sporns2016modular,meunier2010modular}. Such structural diversity has often been associated with enhanced computational flexibility and the coexistence of multiple dynamical regimes within the same network. Although the present study does not attempt to reproduce detailed biological connectivity patterns, the beneficial effect of weakly coupled neuron subsets illustrates how even simple forms of structural diversity can influence the dynamical landscape of recurrent networks.

\NI Another aspect of our findings concerns the frequently discussed hypothesis that optimal information processing occurs near the so-called “edge of chaos.” Early studies in reservoir computing and recurrent neural networks suggested that computational capability can peak near transitions between ordered and chaotic dynamical regimes \cite{bertschinger2004real,legenstein2007edge,boedecker2012information}. Our results confirm that such transition regions can support useful computation in strongly coupled networks. However, they also show that reliance on these narrow parameter regions may not be necessary if regulatory mechanisms are available that actively stabilize the network dynamics. The automatic gain control mechanism explored here effectively replaces the need for precise parameter tuning by maintaining the system in a favorable operating regime over a wide range of structural parameters. From a practical perspective, this suggests that adaptive regulation may represent an alternative design principle for reservoir computers that reduces the sensitivity of the system to the precise choice of network parameters.

\NI Finally, the present results also raise several questions for future research. One important direction concerns the interaction between stabilization mechanisms and the specific computational tasks performed by the reservoir. As discussed above, some tasks may benefit from strongly nonlinear dynamics or even from saturated neuron states that resemble binary logic elements. Other tasks, particularly those requiring faithful representation of input signals over time, may perform best in calmer, quasi-linear regimes. Understanding how different forms of structural heterogeneity or dynamical regulation influence the trade-off between stability and nonlinear computational power remains an open problem. In this context it may also be informative to relate the present observations to studies showing that trained recurrent networks often operate on low-dimensional dynamical manifolds embedded in the high-dimensional state space \cite{sussillo2013opening}.

\NI Another interesting question concerns the scalability of the observed effects. The experiments presented here were performed with relatively small reservoirs in order to allow systematic exploration of the parameter space. In larger networks, collective dynamical phenomena may become even more pronounced, potentially increasing the importance of stabilization mechanisms such as those explored here. It would therefore be valuable to investigate how the proposed structural and dynamical countermeasures perform in much larger reservoirs and in more complex information-processing tasks. In addition, recent work on physical and neuromorphic implementations of reservoir computing suggests that adaptive regulatory mechanisms could play an important role in making such systems robust to parameter variability and environmental fluctuations \cite{tanaka2019recent}.

\NI Taken together, the present study demonstrates that runaway excitation in strongly coupled recurrent neural networks can be mitigated either by introducing subtle structural heterogeneity or by implementing a simple form of global dynamical regulation. Both strategies substantially enlarge the dynamical regime in which reservoir computers remain sensitive to input signals and capable of reliable information processing. These findings suggest that combining structural diversity with adaptive regulation may represent a promising direction for the design of robust reservoir computing systems.

\section{Additional Information}

\subsection{Author contributions}

CM conceived the study, implemented the methods, evaluated the data, and wrote the paper. 
AS discussed the results and acquired funding.
AM and TK discussed the results and provided resources.
PK conceived the study, discussed the results, acquired funding and wrote the paper.

\subsection{Funding}
This work was funded by the Deutsche Forschungsgemeinschaft (DFG, German Research Foundation): grants KR\,5148/3-1 (project number 510395418), KR\,5148/5-1 (project number 542747151), KR\,5148/10-1 (project number 563909707) and GRK\,2839 (project number 468527017) to PK, and grants SCHI\,1482/3-1 (project number 451810794) and SCHI\,1482/6-1 (project number 563909707) to AS.

\subsection{Competing interests statement}
The authors declare no competing interests.

\subsection{Data availability statement} 
The complete data and analysis programs will be made available upon reasonable request.

\subsection{Third party rights}
All material used in the paper are the intellectual property of the authors.


\newpage
\bibliographystyle{unsrt}
\bibliography{references}


\newpage


\newpage

\newpage

\newpage

\end{document}